\documentclass[lettersize,journal]{IEEEtran}

\usepackage{amsmath,amssymb,amsfonts,bm}
\usepackage{array,booktabs,multirow,makecell,longtable,tabularx,adjustbox,ragged2e}
\setcellgapes{3pt}

\usepackage{graphicx}
\usepackage[caption=false,font=normalsize,labelfont=sf,textfont=sf]{subfig}
\usepackage{stfloats}

\usepackage{algorithm,algorithmic}

\usepackage[justification=centering]{caption}
\usepackage[utf8]{inputenc}

\usepackage{cite}
\usepackage{tabularx} % 导言区加入
\usepackage{xcolor}
\usepackage{url}
\usepackage{verbatim}
\usepackage{ragged2e}
\usepackage{changepage}
\usepackage{pifont}

\usepackage[colorlinks=true,linkcolor=blue,citecolor=blue,urlcolor=blue]{hyperref}

\begin{document}

\title{Vision–Language–Action Models Meet World Models: Embodied Agentic AI for Low-Altitude Wireless Networks}

\author{Feibo Jiang, \textit{Senior Member, IEEE}, Li Dong, Lei Mao, Kezhi Wang, \textit{Senior Member, IEEE}, Cunhua Pan, \textit{Senior Member, IEEE}, Dong In Kim, \textit{Life Fellow, IEEE}, and Naofal Al-Dhahir, \textit{Fellow, IEEE}.
	%	\thanks{Feibo Jiang (jiangfb@hunnu.edu.cn) is with Hunan Provincial Key Laboratory of Intelligent Computing and Language Information Processing, Hunan Normal University, Changsha, China.}
}

\maketitle

\begin{abstract}
Low-Altitude Wireless Networks (LAWNs), composed of Unmanned Aerial Vehicles (UAVs) and other aerial platforms, provide integrated perception, communication, and computation services in low-altitude airspace. However, deploying large generative models in this domain faces three major challenges: 1) Limited embodied action mapping; 2) Inadequate physical environment modeling; 3) Insufficient closed-loop optimization. To address these challenges, this study proposes an Embodied Agentic UAV framework. Centered on a Vision–Language–Action (VLA) model as the execution core, the framework establishes an end-to-end embodied decision-making pipeline from multimodal environmental perception to continuous control generation. In addition, a World Model (WM) is introduced to capture the coupling between UAV actions and environmental state evolution, thereby supporting environment prediction, policy verification, and dynamic optimization. Furthermore, memory and reflection mechanisms are incorporated to form an adaptive closed-loop optimization paradigm of decision, execution, evaluation, and update, thereby enhancing the system’s autonomous decision-making capability and continual evolution ability in complex dynamic environments. Experimental results validate its effectiveness in enabling robust, predictive, and sustainable autonomous control in LAWNs.
\end{abstract}


\section{Introduction}
Low-Altitude Wireless Networks (LAWNs) refer to integrated air–ground communication systems operating below a few hundred meters, where aerial platforms such as Unmanned Aerial Vehicles (UAVs) provide communication, sensing, and computing services. With rapid deployment and flexible reconfigurability, LAWNs enable on-demand connectivity in infrastructure-limited or bursty-traffic scenarios, and are becoming a key enabler of the low-altitude economy and air–ground integrated systems. However, the 3D mobility of aerial nodes leads to rapidly evolving network topologies \cite{jin2026advancing}, while air–ground channels are highly time-varying and vulnerable to blockage and interference. In addition, communication, sensing, and control are tightly coupled and constrained by flight safety, onboard computation, and limited battery capacity, necessitating autonomous and intelligent decision-making in dynamic and uncertain environments\cite{8770066}.

Recent advances in Large Language Models (LLMs) and Vision Language Models (VLMs) have demonstrated strong semantic understanding and cross-modal reasoning capabilities \cite{11370829}. Centered on natural language, LLMs support high-level intent understanding, task decomposition, and constraint representation, providing goal-oriented reasoning and decision support for complex systems. VLMs further align linguistic semantics with environmental perception, enabling semantically structured scene representations that, fused with wireless physical states, yield a more comprehensive basis for decision-making and improve interpretability. In LAWNs, where network topologies and channel conditions change rapidly, LLM/VLM-based cognitive modules offer clear advantages in task alignment and explainable reasoning. However, they typically lack explicit modeling of action–environment dynamics, leading to the following challenges in LAWNs:

\subsubsection{Limited Embodied Action Mapping}
Existing LLM/VLM-based approaches in LAWNs mainly focus on semantic understanding and high-level strategy generation, but lack an end-to-end mechanism that maps task objectives and multimodal observations directly to executable control actions \cite{11164279}. In LAWNs, online decisions must satisfy flight dynamics, collision avoidance, and multiple operational constraints simultaneously. Without an embodied action interface, decision outputs cannot be expressed in standardized parametric forms, nor reliably projected onto feasible solution spaces or validated under strict constraints. This limitation significantly reduces deployability, controllability, and closed-loop stability in practical systems.

\subsubsection{Inadequate Physical Environment Modeling}
Online decision-making in LAWNs relies on continuous interaction with the physical environment and accurate modeling of its evolving dynamics. Although VLMs provide visual–semantic scene understanding, they lack unified representations and cross-modal alignment to jointly model wireless physical quantities and network states \cite{10929033}. As a result, systems struggle to associate visual semantics with key factors such as blockage, multipath effects, energy constraints, and dynamic traffic loads during trajectory optimization and network evolution. This deficiency weakens predictability, generalization, and decision reliability in highly dynamic environments.

\subsubsection{Insufficient Closed-Loop Optimization}
LAWNs require continuous decision-making in dynamic environments, where trajectory and resource allocation decisions have long-term impacts on energy consumption and task performance. However, existing LLM/VLM-based methods are largely limited to single-shot or stage-wise decisions, lacking mechanisms for continuous optimization across ``planning–execution–feedback–optimization" cycles \cite{9967016}. In addition, their decision processes often lack interpretability and consistent risk evaluation, hindering stable long-term optimization and reliable system operation.

\setcounter{subsubsection}{0}

To address these challenges, this paper integrates a Vision–Language–Action (VLA) model with a World Model (WM) and proposes an Embodied Agentic UAV framework for LAWNs. The main contributions are summarized as follows:

\subsubsection{VLA Perspective}
We propose an executable control module tailored for LAWNs that unifies multimodal perception, language understanding, and action generation within an end-to-end control framework. The model first encodes multimodal perceptual information collected from the LAWN environment into a shared state representation. Task objectives are then injected as conditional signals, enabling high-level semantic intentions to directly influence UAV action generation. Instead of producing abstract policies or discrete decisions, the VLA module directly decodes continuous low-dimensional action vectors describing UAV maneuvers, aligned with flight-control interfaces. This design establishes a real-time control loop driven by semantic goals.

\subsubsection{World Model Perspective}
We further introduce a WM designed for LAWNs. Built upon multimodal continuous observations, the model first learns a visual world representation that preserves long-term video consistency and spatial memory. Flight actions are explicitly incorporated as driving factors to model how UAV control inputs influence environmental state evolution. By learning the continuous dynamics of world states conditioned on UAV actions, the WM enables persistent internal simulation and trend prediction of UAV flight processes. This results in an interactive and predictive environmental abstraction that provides a stable and generalizable physical-world representation for offline policy learning and online optimization.

\subsubsection{Agentic System Perspective}
We construct an embodied Agentic system for LAWNs. The VLA module serves as the action agent, generating executable actions with the support of the memory module. The memory module retrieves effective strategies and risk boundaries from similar historical scenarios to guide current decision-making. The reflection module evaluates execution outcomes with predictive support from the WM. Subsequently, key experiences and causal explanations are written back into memory in a structured form, with their validity and confidence continuously updated. Through the closed-loop process of “VLA decision → execution → WM evaluation → memory update,” the system achieves continuous self-adaptive optimization.

The remainder of this paper is organized as follows. Section II reviews the evolution from perception to action. Section III presents the integration of language, vision, and world modeling. Section IV discusses the progression from AI agents to embodied Agentic AI. Section V introduces the proposed Embodied Agentic UAV framework. Section VI provides experimental results. Section VII discusses open issues, and Section VIII concludes the paper.

\section{From VLM to VLN to VLA: The Evolution from Perception to Action}
In the evolutionary trajectory of large models, VLMs, Vision–Language Navigation (VLN) models, and VLA models represent successive capability upgrades from semantic understanding to spatial decision-making and ultimately to executable action generation, forming a technological progression toward embodied intelligence.

\subsection{VLMs for Multimodal Semantic Perception}
Within LAWNs, VLMs serve as the core component for multimodal semantic perception and situational understanding. Their main role is to align UAV-perspective images or videos, geographic environmental information, and task descriptions into a unified semantic space. Through such cross-modal alignment, VLMs generate structured and interpretable environmental representations, enabling the network to characterize scene composition, task requirements, and potential constraints semantically. These representations provide cognitive-level inputs for deployment decisions and resource management. More importantly, they improve scene understanding and support knowledge transfer across diverse operational scenarios.

\subsection{VLN for Goal-Driven Spatial Navigation}
Building upon semantic understanding, VLN introduces spatial decision-making by jointly mapping task objectives and visual or map-based perceptions into navigation policies\cite{liu2026dronenav}. In LAWNs, this capability naturally supports UAV placement and trajectory planning, translating high-level task requirements into executable navigation strategies while allowing dynamic adjustments during mission execution. Compared with VLMs, which focus on environmental understanding, VLN transforms semantic task descriptions into spatially executable flight strategies. In addition, it can incorporate communication performance and resource consumption into navigation decisions, enabling deployment strategies that jointly consider mission objectives, safety, coverage quality, link reliability, and energy efficiency.

\subsection{VLA for End-to-End Action Generation}
VLA further extends the paradigm from understanding and navigation to fully executable end-to-end action generation, enabling the direct production of UAV motion control commands. By jointly modeling visual perception, task objectives, and action generation within a unified framework, VLA addresses the tightly coupled communication, control, and safety requirements in LAWNs’ online decision-making. High-level semantic goals can thus be translated into executable control signals, improving real-time responsiveness and closed-loop control efficiency. Compared with VLN, which mainly focuses on position- and trajectory-level decisions, VLA provides a more complete action space, outputs more aligned with execution interfaces, and stronger closed-loop control capability. These advantages make it a key foundation for autonomous control and optimization in LAWNs.

In summary, VLMs, VLN, and VLA represent a progressive evolution of intelligent capabilities in LAWNs, from semantic understanding to executable action. VLMs provide semantic perception and situational awareness, VLN enables goal-oriented navigation and deployment, and VLA further supports executable control and online closed-loop optimization. Together, they form an embodied intelligence pipeline that lays the foundation for autonomous and robust LAWNs.

\begin{figure}[htbp]
	\centering
	\includegraphics[width=8.5cm]{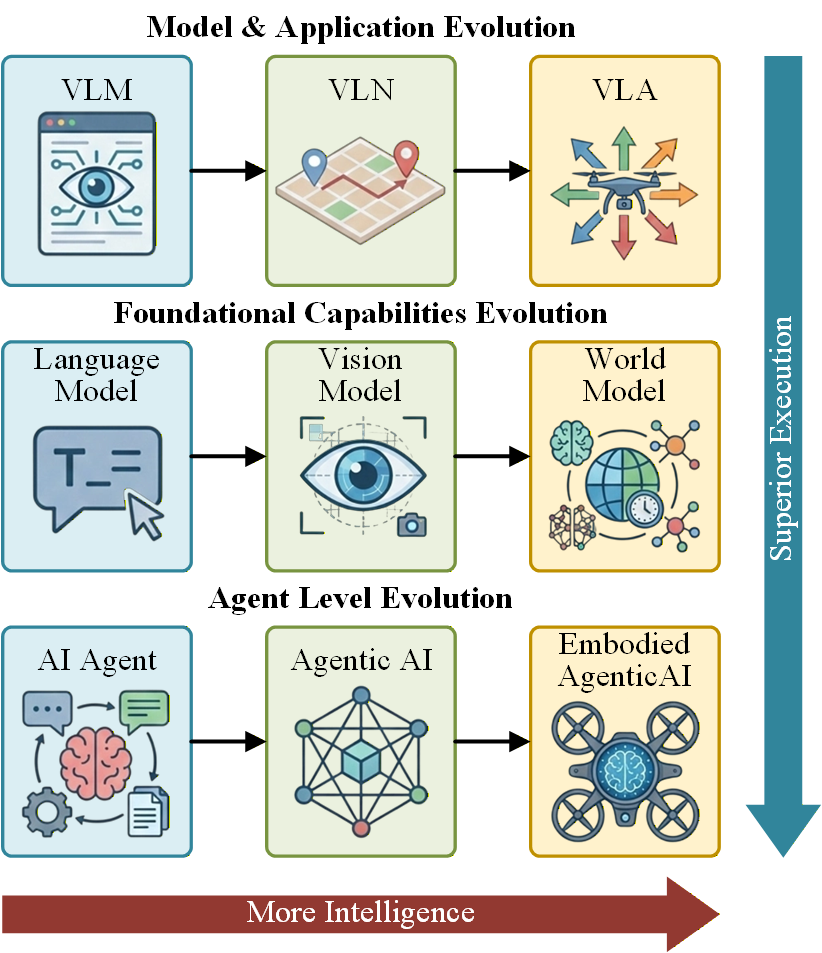}
	\caption{A Development Roadmap Toward Embodied Agentic AI for LAWNs.}
	\label{fig:fig1}
\end{figure}

\section{From Language and Vision to World Models: Toward Unified Intelligence}
In the intelligent evolution of LAWNs, Language Models (LMs), Vision Models (VMs), and WMs correspond respectively to three fundamental capabilities: semantic reasoning, environmental perception, and dynamic prediction. Together, they form a progressive framework that advances network intelligence from static cognition toward embodied autonomy.

\subsection{Language Models: Task Understanding and Semantic Representation}
In LAWNs, the primary value of LMs lies in their natural language understanding and abstract reasoning capabilities, enabling networks to interpret high-level task descriptions and transform them into structured intents, operational constraints, and optimization objectives. Compared with conventional network control methods based on predefined rules or explicit analytical models, LMs leverage knowledge and strategy patterns learned from large-scale corpora to support task decomposition, priority organization, and decision logic formulation in complex mission settings. As a result, network intelligence evolves from parameter-level adjustment toward goal-oriented semantic decision-making. However, these capabilities still remain largely within symbolic reasoning and static decision processes, lacking direct perception of environmental dynamics and explicit modeling of action consequences.

\subsection{Vision Models: Physical-World Perception and Scene Understanding}
Building upon the semantic reasoning capability of LMs, VMs introduce embodied perception of the physical environment into LAWNs. Through visual observations, aerial nodes can acquire environmental structures and operational contexts, compensating for the state information limitations of approaches relying only on wireless measurements or prior rules. Visual perception enables the construction of physically consistent state representations, providing more reliable environmental references for signal coverage optimization, trajectory adjustment, and safety control. Nevertheless, VMs mainly focus on observation interpretation; their outputs primarily characterize the current scene rather than modeling how the environment evolves over time under agent actions.

\subsection{World Models: Physical-World Dynamics and Evolution Prediction}
WMs further extend network intelligence from perception and understanding to prediction and planning. Their main objective is to learn the evolution dynamics of environmental and system states conditioned on agent actions, particularly through physics-consistent modeling of state transitions, thereby forming internal representations suitable for multi-step reasoning and simulation. In LAWNs, this capability enables the system to characterize how UAV actions and resource allocation decisions affect subsequent physical environments and network states, providing essential support for rolling planning, risk assessment, and long-term optimization. Unlike LMs and VMs, which mainly depend on observational information, WMs emphasize continuous modeling of physical processes and state evolution, promoting LAWNs toward a model-driven paradigm of autonomous control.

In summary, LMs, VMs, and WMs correspond to three core capabilities in LAWNs: semantic understanding, environmental perception, and dynamic prediction.
LMs interpret task objectives and form semantic representations, VMs capture physical environments and scene states, and WMs model physical-world dynamics to support prediction and planning. Together, they establish a pathway from understanding and perception to dynamic prediction, laying the foundation for autonomous and verifiable LAWNs.
\section{From AI Agents to Agentic AI to Embodied Agentic AI: The Evolution from Virtual to Physical Agents}
To achieve efficient real-time control and long-term optimization, LAWNs require a more advanced intelligent architecture that extends beyond high-level task understanding and decision reasoning to incorporate environmental perception, interaction, and self-optimization. Hence, AI Agents, Agentic AI, and Embodied Agentic AI form a progressive framework from task cognition to autonomous control, driving LAWNs toward greater autonomy, intelligence, and interpretability.

\subsection{AI Agents: Task-Oriented Intelligence}
In LAWNs, AI Agents are typically centered on LLMs and represent the initial stage of network intelligence. Their main capabilities include task understanding, intent interpretation, strategy generation, and optimization. Such agents can receive natural-language instructions and translate them into structured network intents, operational constraints, and decision plans. By leveraging knowledge reasoning and learned experience patterns, they decompose complex tasks, generate high-level control recommendations, and achieve limited online adaptation through feedback. However, these agents remain disembodied cognitive entities, as they neither directly perceive the physical environment nor explicitly model how their decisions affect flight trajectories, network topology evolution, or safety conditions.

\subsection{Agentic AI: Collaborative Multi-Agent Intelligence}
With the introduction of multi-agent systems, low-altitude network intelligence evolves from static cognitive agents to Agentic AI, characterized by continuous perception, feedback, and adaptive strategy adjustment through collaboration\cite{11370176}. In such systems, some agents specialize in real-time environmental perception using multimodal sensing, such as aerial vision and radar, to capture environmental maps, obstacle distributions, traffic conditions, and risk regions. Other agents focus on planning, decision-making, and reflective reasoning based on perceived information. Through this collaborative structure, Agentic AI continuously couples task objectives with dynamic environmental states, enabling flexible and context-aware decision updates. Nevertheless, it still mainly follows a perception–response paradigm and lacks deep modeling of long-term evolution and causal effects.

\subsection{Embodied Agentic AI: World-Model-Driven Physical Agents}
When VLA models and WMs are integrated into the Agentic AI framework, LAWNs further evolve into Embodied Agentic AI, forming a closed-loop autonomy cycle of perception, decision, action, feedback, and optimization \cite{11449255}. The WM learns and simulates latent network states and dynamic mechanisms, capturing causal relationships among UAV trajectories, channel evolution, energy consumption, and task utility. As a result, agents can not only perceive the environment but also conduct multi-step counterfactual reasoning in simulated environments, predicting the long-term impact of alternative actions on network performance and safety risks. Meanwhile, the VLA module serves as the action layer, directly generating actionable control commands aligned with physical interfaces. Through this integration, network control advances from reactive response to predictive, verifiable, and continuously optimizable autonomous behavior.

In summary, the intelligence of LAWNs evolves from language-driven task cognition to collaborative decision-making and ultimately to WM-driven autonomous control. The LM-based agent provides the foundation for task understanding and decision representation, multi-agent collaboration enables adaptive behavior, and the integration of WMs with VLA allows agents to anticipate future outcomes and execute actions accordingly. This progression outlines a clear pathway toward long-term autonomous and robust LAWNs.

\section{Embodied Agentic UAV Framework}
\begin{figure*}[htbp]
	\centering
	\includegraphics[width=16cm]{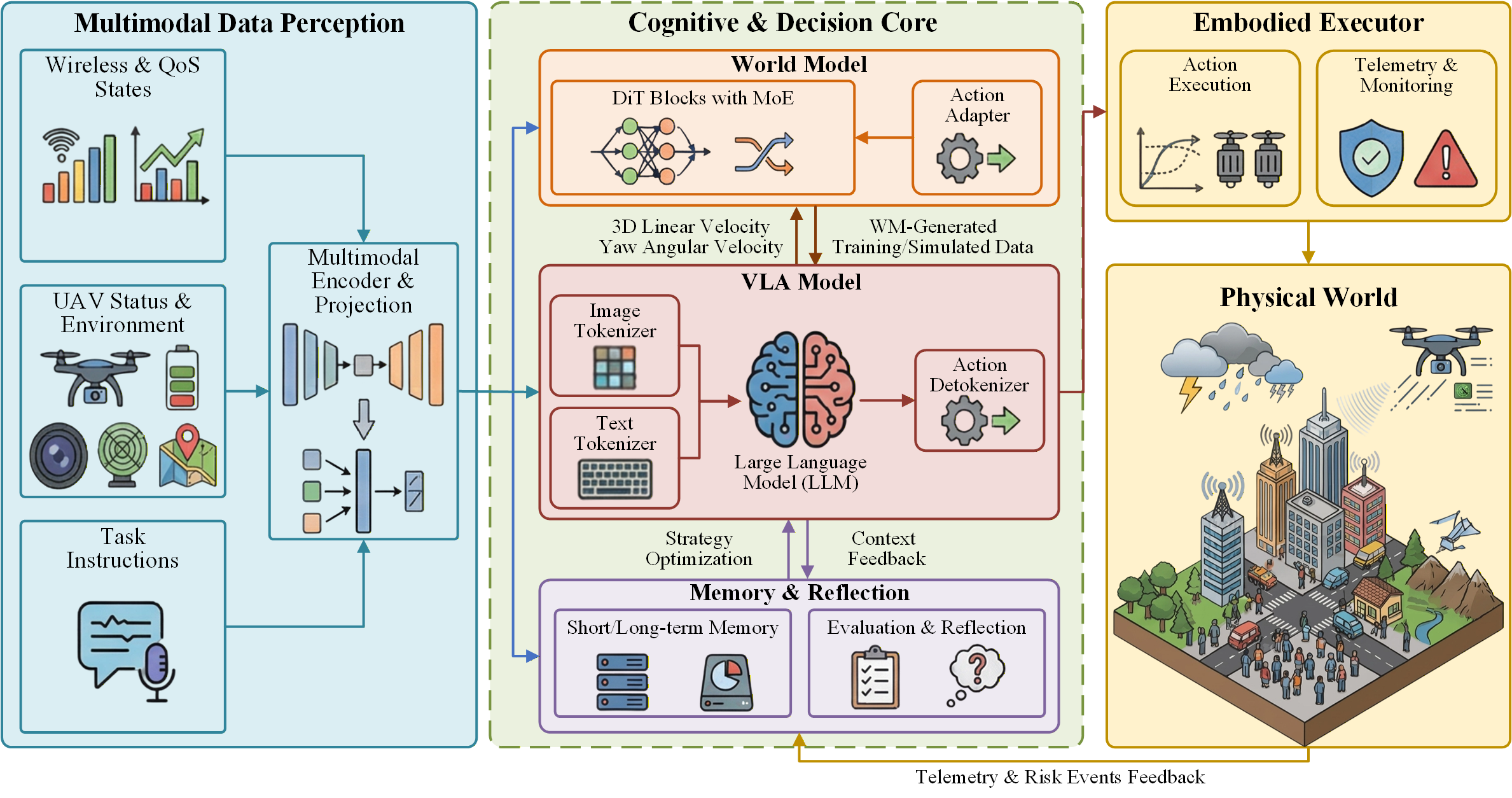}
	\caption{System Architecture of the Embodied Agentic UAV Framework.}
	\label{fig:fig2}
\end{figure*}

As illustrated in Fig. \ref{fig:fig2}, the Embodied Agentic UAV framework consists of five key components: multimodal data perception, the VLA module, the WM, memory and reflection, and the embodied executor. Together, these components support autonomous perception, decision-making, and closed-loop control in LAWNs.
\subsection{Multimodal Data Perception}
The framework first establishes a multimodal perception layer tailored for LAWNs to provide unified and reliable environmental and network-state inputs.

\subsubsection{Data Acquisition}
This layer collects heterogeneous information from both the physical environment and network operation. The acquired data include wireless measurement information (e.g., channel quality statistics, interference levels, and queue-length indicators), UAV status data (e.g., flight attitude, velocity, altitude, battery level, and onboard device temperature), environmental perception data (e.g., camera images, radar signals, and map information), as well as Quality-of-Service (QoS) requirements. Together, these inputs provide comprehensive situational awareness and form the foundation for subsequent decision-making and control processes.

\subsubsection{Data Preprocessing}
To ensure consistency across heterogeneous sensing sources, the system performs data alignment, denoising, missing-value completion, and anomaly detection, enabling accurate temporal synchronization and removal of noise and redundant information. The processed multimodal inputs are then encoded through a multimodal encoder to produce a unified state representation. Subsequently, a projection module performs feature fusion, transforming raw multimodal observations into high-quality state features suitable for downstream decision-making and optimization.

\subsection{VLA}
The VLA model integrates visual encoding with language reasoning to enable autonomous UAV navigation and real-time decision-making in complex environments.

\subsubsection{Model Architecture}
\begin{figure}[htbp]
	\centering
	\includegraphics[width=8cm]{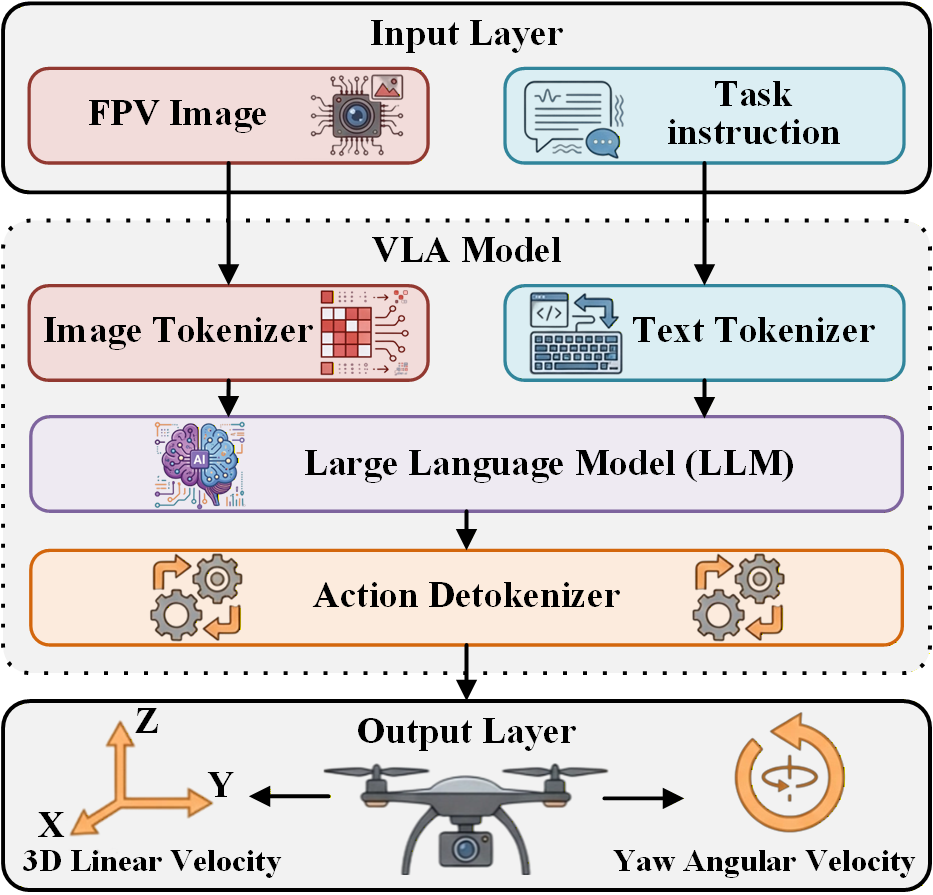}
	\caption{End-to-End VLA Pipeline from Multimodal Input to Action Output.}
	\label{fig:fig3}
\end{figure}

As illustrated in Fig. \ref{fig:fig3}, the VLA model follows an end-to-end architecture from multimodal input to action output. Specifically, FPV images captured by the UAV camera and natural-language task instructions are first fed into the input layer. The visual and textual inputs are then processed by an image tokenizer and a text tokenizer, respectively, and mapped into a shared representation space for joint reasoning in the LLM. Based on the fused multimodal representations, an action detokenizer generates executable control outputs, including three-dimensional linear velocity and yaw angular velocity. Compared with conventional waypoint-based planning methods, this architecture directly produces low-level control commands in real time, enabling continuous adaptation to dynamic environments and efficient UAV trajectory control.
\subsubsection{Training and Deployment}
During training, the VLA model uses both real flight data and WM-generated simulated data, including FPV images, video sequences, and corresponding task instructions. The training process begins with supervised learning to establish the mapping from multimodal inputs to action outputs, followed by fine-tuning and reinforcement learning to improve control performance under dynamic environments. To enhance training efficiency and inference speed, LoRA is adopted for parameter-efficient fine-tuning, while quantization is applied during deployment to reduce computational overhead \cite{wang2026vla}. These optimizations enable efficient real-time decision-making and control under limited onboard resources, allowing UAVs to respond rapidly in complex environments.

\subsection{World Model}
The proposed WM serves as a predictive environment model for LAWNs, enabling interactive simulation of UAV flight processes and physical-world evolution under action and instruction conditions.

\subsubsection{Model Architecture}
\begin{figure}[htbp]
	\centering
	\includegraphics[width=8cm]{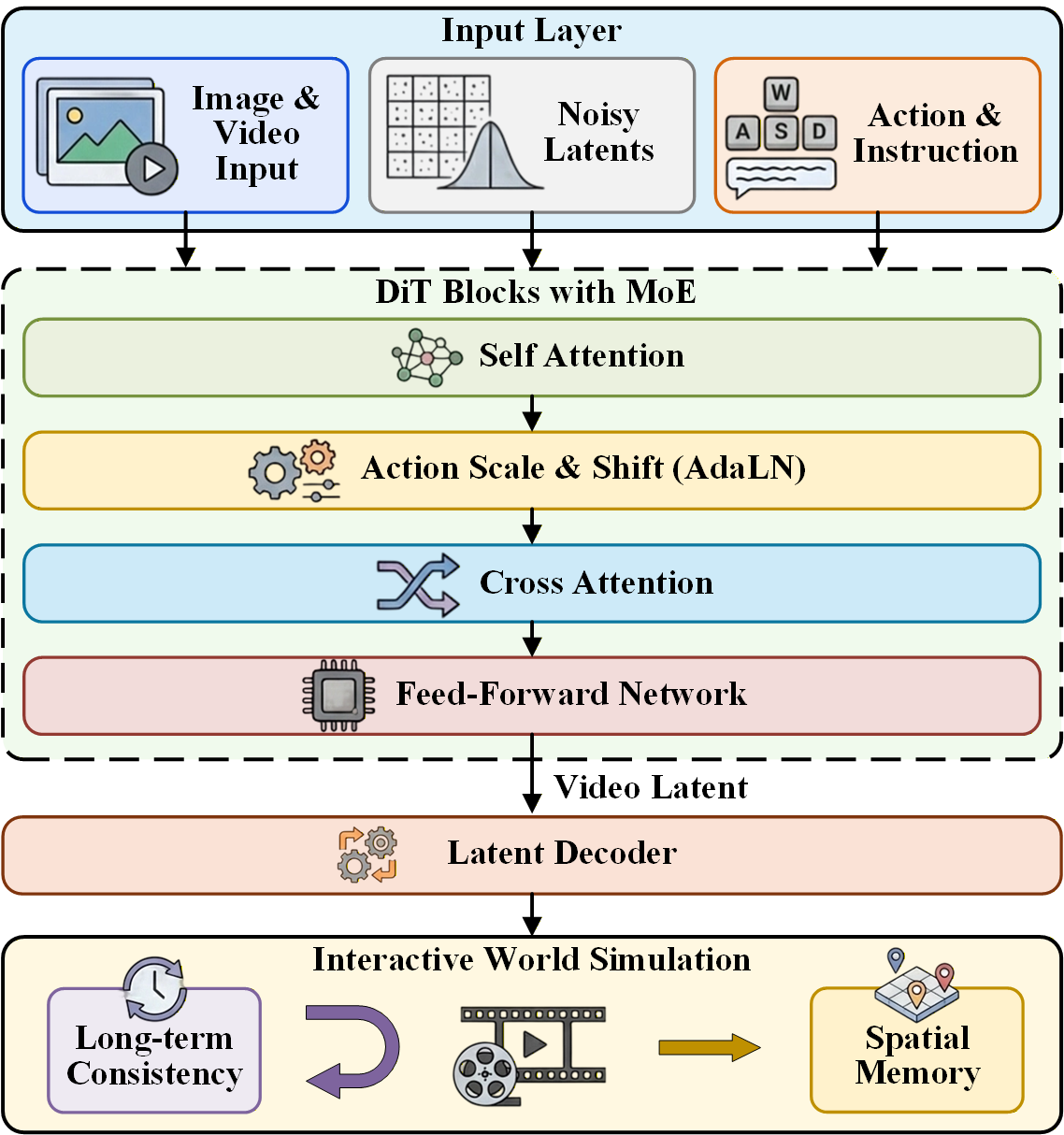}
	\caption{Architecture of the DiT-Based World Model for Interactive World Simulation.}
	\label{fig:fig4}
\end{figure}

The WM is built upon a Diffusion Transformer (DiT) backbone equipped with a Mixture-of-Experts (MoE) design to enhance generative capability while maintaining computational efficiency. As illustrated in Fig. \ref{fig:fig4}, the input layer consists of image and video observations, noisy latent variables, and action-instruction conditions. These inputs are jointly processed by stacked DiT blocks, in which self-attention captures global dependencies within the latent representations, action-conditioned adaptive layer normalization injects control information through scale-and-shift modulation, cross-attention aligns latent states with external conditions, and a feed-forward network further refines the hidden features. The DiT backbone outputs video latents, which are then converted by a latent decoder into future simulated observations. With this design, the world model supports interactive world simulation with long-term temporal consistency and spatial memory, thereby capturing how agent actions influence future environmental evolution.
\subsubsection{Training Strategy}
The WM is trained using a multi-stage progressive learning strategy. First, a foundation DiT model pretrained on large-scale datasets is adopted as a prior, ensuring high-fidelity generation of static physical environments. Next, the backbone network is largely frozen while the Action Adapter is fine-tuned, enabling progressive expansion from short-horizon generation to long-horizon state evolution modeling. This gradual training process equips the model with long-term temporal consistency and action-conditioned responsiveness. Finally, the framework focuses on improving real-time performance and stability. Distribution Matching Distillation (DMD) and adversarial optimization, implemented via an auxiliary discriminator, are employed to compress diffusion sampling steps, thereby reducing inference latency and enabling real-time interactive simulation \cite{yin2024one}.

\subsection{Memory and Reflection}
\subsubsection{Memory-Augmented Module}
The memory module enhances decision efficiency and quality through a hierarchical design consisting of short-term working memory and long-term experiential memory. Short-term memory provides context-aware references from similar recent scenarios, enabling the VLA agent to rapidly adapt through in-context reasoning and improving response efficiency. Long-term memory stores historical task experiences and refinement strategies, including successful policies, failure cases, and domain knowledge across representative scenarios. These stored experiences serve as references for evaluation and reflection processes, supporting continual improvement of decision quality.

\subsubsection{Evaluation and Reflection Module}
The evaluation and reflection module is responsible for real-time assessment of execution outcomes and generation of interpretable decision rationales. After action execution, the reflection module evaluates performance based on key metrics such as task completion status, latency, and energy consumption. Leveraging the WM, the system performs predictive analysis to derive improved strategies and produces structured explanations, including why a specific action was taken, expected benefits, and potential risks. These feedback signals provide verifiable evidence for decision-making and are written back into long-term memory in structured form. Through continuous accumulation and updating of experience confidence, the system progressively refines its policies, enabling increasingly stable and robust control and decision-making during long-term operation.

\subsection{Embodied Executor}
The embodied executor provides deterministic-latency and high-reliability execution capabilities for real UAV networks while returning operational states and risk events to support upper-layer closed-loop optimization.

\subsubsection{Action Execution Module}
This module standardizes and encapsulates the action vectors generated by the VLA model, performing protocol adaptation to map them into control command sequences directly interpretable by the flight control module. At the flight-control module, a trajectory-tracking controller converts high-level commands into motor thrust and attitude control signals, enabling closed-loop regulation of roll, pitch, yaw, and altitude stabilization. Operating under strict control frequencies, the UAV maintains stable flight while satisfying the real-time responsiveness, repeatability, and safety requirements of low-altitude operations.

\subsubsection{Telemetry and Monitoring Module}
The module performs two primary functions. First, it collects and transmits critical telemetry data—including attitude, velocity, altitude, remaining energy, device temperature, link quality, throughput, and latency—to support evaluation and reflection processes. Second, it provides real-time safety monitoring and arbitration by continuously checking no-fly zones, collision risks, and velocity or altitude constraints. When safety violations are detected, commands are automatically degraded or rewritten, and fallback policies are triggered. This mechanism ensures stable hardware operation under strict safety constraints while simultaneously providing reliable and auditable real-world data for VLA feedback learning and WM calibration.

\section{Experimental Results}
\subsection{Experimental Setup}
We constructed a simulation environment to evaluate the proposed Embodied Agentic UAV framework. The VLA module was implemented based on the OpenVLA-7B model and fine-tuned using LoRA \cite{kim2025openvla}. The WM adopts the open-source LingBot-World architecture, which incorporates both high-noise and low-noise expert modules, with a total parameter scale of approximately 28B \cite{team2026advancing}.

 \textcolor{black}{The VLA fine-tuning dataset is based on the RaceVLA dataset, which contains UAV First-Person-View (FPV) images paired with corresponding four-dimensional action vectors and natural-language task descriptions \cite{serpiva2025racevla}. Following the RaceVLA setting, the UAV dynamics and channel model were adopted in our simulation. The dataset was split into training and testing subsets with a ratio of 70\% and 30\%, respectively. Each task was evaluated over 100 independent trials, and the results were averaged across five random seeds.} To improve data diversity and generalization capability, an additional 50\% of synthetic data was generated using the WM and incorporated into training. The memory module was implemented using Milvus, while the reflection module was powered by DeepSeek R1. Experiments were conducted on a hardware platform equipped with an Intel Xeon Gold CPU (2.6 GHz) and two NVIDIA A800 GPUs (80 GB VRAM each).
\begin{comment}
	内容...

\begin{table}[t]
\centering
\renewcommand{\arraystretch}{1.15}
\setlength{\tabcolsep}{6pt} % 调整列间距

\begin{tabular}{|
>{\centering\arraybackslash}p{1cm}|
>{\centering\arraybackslash}p{1.3cm}|
>{\centering\arraybackslash}p{1.3cm}|
>{\centering\arraybackslash}p{1.3cm}|
>{\centering\arraybackslash}p{1cm}|}
\hline
\textbf{Model} & \textbf{Task-1 SR (\%)} & \textbf{Task-2 SR (\%)} & \textbf{Task-3 SR (\%)} & \textbf{Avg. SR (\%)} \\
\hline
RT-2-X & 53.45 & 37.57 & 51.67 & 47.56 \\
\hline
OpenVLA & 61.28 & 35.46 & 85.34 & 60.69 \\
\hline
\textbf{Ours} & \textbf{74.67} & \textbf{44.33} & \textbf{87.65} & \textbf{68.22} \\
\hline
\end{tabular}
\caption{Success Rate (SR) comparison across different tasks and models. Avg. SR is the mean over Task-1, Task-2, and Task-3.}
\label{tab:SR_comparison}
\end{table}
\end{comment}
\begin{figure}[htbp]
	\centering
	\includegraphics[width=9cm]{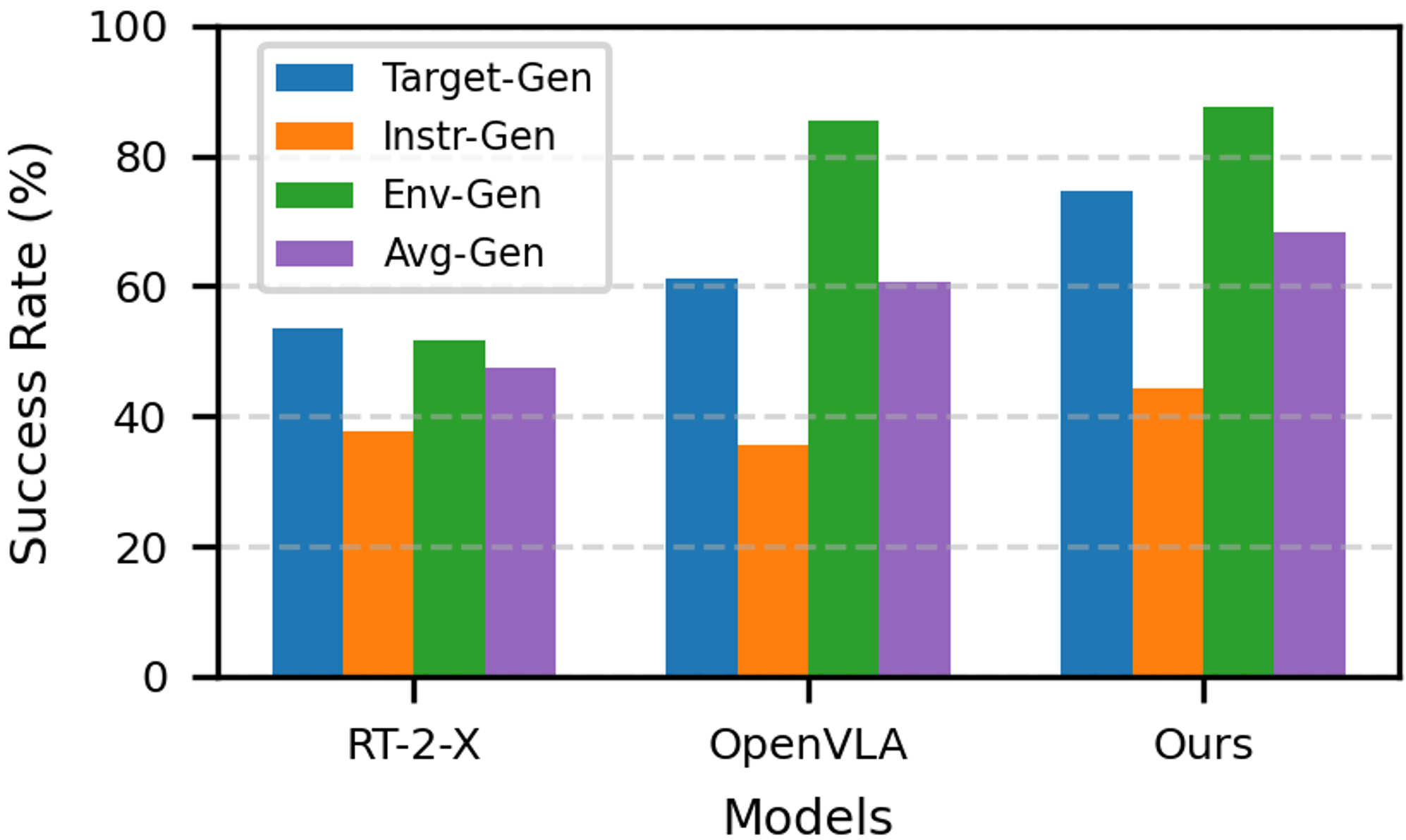}
	\caption{SR Comparison of VLA Models under Three Generalization Tasks.}
	\label{fig:fig5}
\end{figure}

%We first evaluate the ability of the VLA module to transform natural-language instructions and multimodal perception inputs into executable flight actions, focusing on both accuracy and robustness. The task objective requires the UAV to reach a single target region without collision while satisfying the given instruction. The proposed VLA model is compared against RT-2-X [X] and OpenVLA [X]. The evaluation metric is Success Rate (SR), defined as the percentage of trials in which the UAV successfully completes the task without collision while fulfilling the instruction requirements. Performance is evaluated under three task scenarios: (1) Target-Gen Task: SR under previously unseen target locations; (2) Instr-Gen Task: SR under previously unseen language instructions; (3) Env-Gen Task: SR under previously unseen environments and backgrounds. As illustrated in Fig. \ref{fig:fig5},  our VLA model achieves the highest SR across all tasks, demonstrating superior capability in understanding UAV task instructions and producing effective end-to-end action control. These results indicate that our  VLA provides significant advantages in instruction grounding and real-time control generation compared with existing approaches.

We first evaluate the ability of the VLA module to transform natural-language instructions and multimodal perception inputs into executable flight actions, focusing on both accuracy and robustness. The task requires the UAV to reach a single target region without collision while satisfying the given instruction. The proposed VLA model is compared with RT-2-X \cite{o2024open} and OpenVLA \cite{kim2025openvla}. Success Rate (SR) is used as the evaluation metric, defined as the percentage of trials in which the UAV completes the task collision-free while fulfilling the instruction requirements. Performance is evaluated under three generalization tasks: (1) Target-Gen Task: SR under previously unseen target locations; (2) Instr-Gen Task: SR under previously unseen language instructions; (3) Env-Gen Task: SR under previously unseen environments and backgrounds. As shown in Fig. \ref{fig:fig5}, our VLA model achieves the highest SR across all three generalization tasks, demonstrating clear advantages in instruction grounding and real-time action generation over existing baselines.

\begin{figure}[htbp]
	\centering
	\includegraphics[width=9cm]{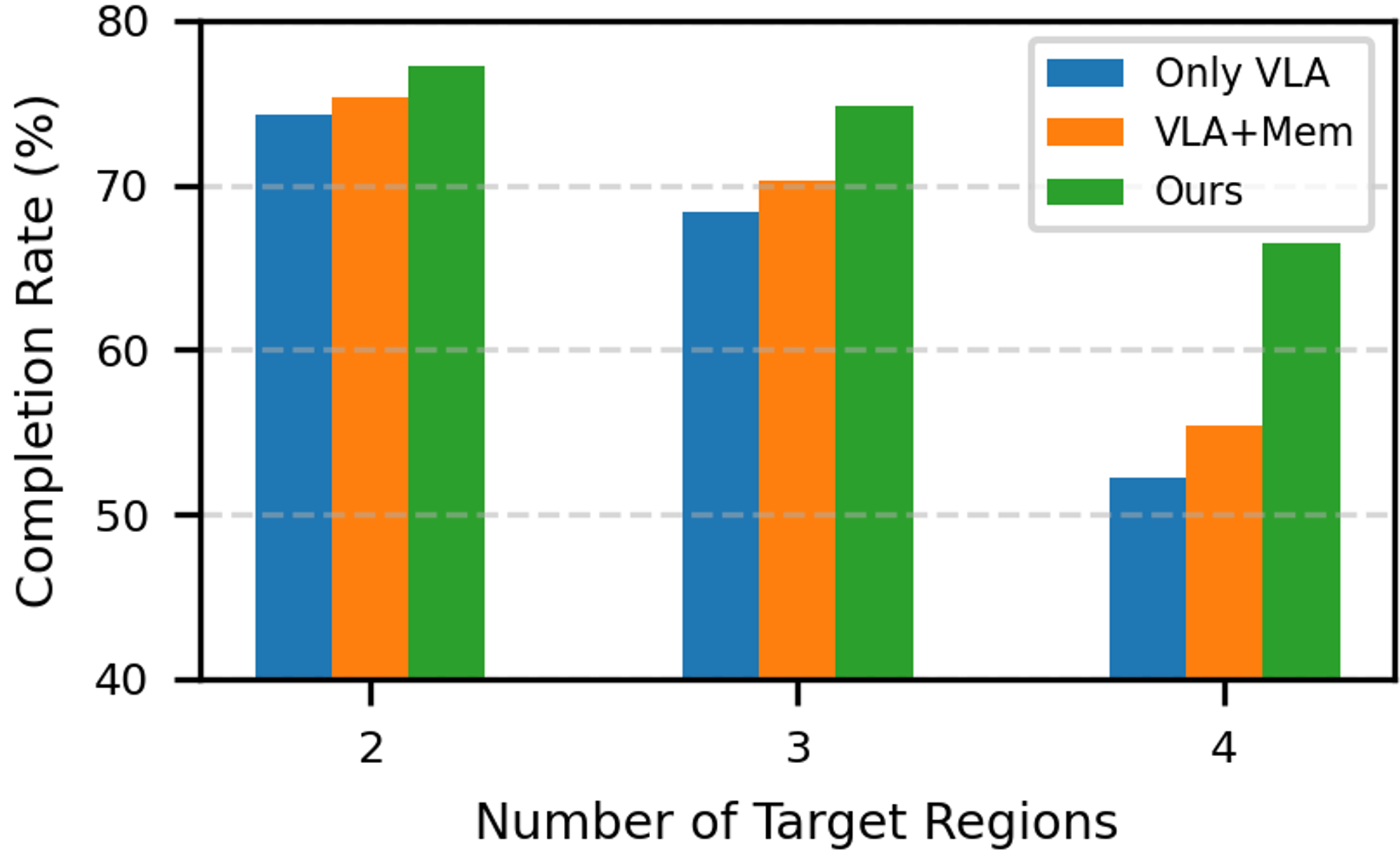}
	\caption{Typical Application Scenarios of Aerial Agentic AI.}
	\label{fig:fig6}
\end{figure}

\begin{comment}
\begin{table}[t]
\centering
\renewcommand{\arraystretch}{1.15}
\setlength{\tabcolsep}{6pt} % 调整列间距

\begin{tabular}{|
>{\centering\arraybackslash}p{2.5cm}|
>{\centering\arraybackslash}p{1.5cm}|
>{\centering\arraybackslash}p{1.5cm}|
>{\centering\arraybackslash}p{1.5cm}|}
\hline
\textbf{Model} & \textbf{2 Targets} & \textbf{3 Targets} & \textbf{4 Targets} \\
\hline
VLA & 74.31 & 68.43 & 52.27 \\
\hline
VLA + Memory & 75.35 & 70.32 & 55.36 \\
\hline
\textbf{Ours} & \textbf{77.26} & \textbf{74.86} & \textbf{66.48} \\
\hline
\end{tabular}
\caption{Task completion rate for different numbers of target regions.}
\label{tab:completion_rate}
\end{table}
\end{comment}
We further evaluate the performance of the complete Embodied Agentic UAV system, which integrates the WM and memory–reflection mechanisms. In this setting, the UAV is required to sequentially reach multiple target regions without collision. The compared methods include: (1) VLA only; (2) VLA with memory and reflection; (3) Ours (VLA + Memory + WM). The evaluation metric is Completion Rate (CR), defined as the ratio between the number of successfully reached target regions and the total number of target regions under collision-free conditions. As shown in Fig. \ref{fig:fig6}, our method achieves the highest CR among all baselines. This improvement primarily stems from the synergy between the WM and the memory mechanism. Specifically, the WM can internally simulate the potential outcomes of alternative action sequences, thereby providing verifiable foresight for policy adjustment, while the memory module retrieves relevant historical strategies and failure cases from similar scenarios to guide decision-making.

\section{Open Issues}
\subsubsection{VLA Inference Latency}
In LAWNs, the VLA module must translate high-level task instructions and multimodal observations into continuous control commands in real time. However, current VLA implementations often rely on autoregressive inference, leading to latency fluctuations caused by context length and visual encoding overhead. A key challenge is how to achieve low-latency and low-jitter execution without significantly degrading decision quality. \textcolor{black}{Possible directions include parallel or non-autoregressive action prediction, edge-cloud collaborative inference, and model distillation and quantization for stable real-time UAV control.}

\subsubsection{World Model Accumulated Error}
The WM supports rolling prediction and planning by learning environmental and system dynamics, but its predictive process in LAWNs is vulnerable to cumulative errors. The nonlinear and time-varying nature of UAV motion and energy consumption causes approximation errors to grow over multi-step prediction. In addition, unmodeled factors such as dynamic occlusions and external disturbances further increase deviation, causing long-horizon predictions to drift from actual trajectories. How to explicitly model and mitigate such accumulated errors remains an important open problem.

\subsubsection{Agentic Learning Stability}
For long-term deployment, UAV agents must continually incorporate new experiences and update policies in non-stationary environments. However, online updating may introduce catastrophic forgetting and distributional shift, making stable performance across new scenarios difficult to maintain. Key open issues include how to coordinate short-term contextual memory with long-term experiential memory for controlled writing, effective retrieval, and dynamic updating, as well as how to build stable continual learning pipelines that preserve robust generalization and long-term performance under cross-task distribution shifts.

\subsubsection{Delay-Doppler Domain Representation} \textcolor{black}{In LAWNs, high UAV mobility causes rapidly varying channel and network states, requiring frequent updates to avoid outdated state information. This increases the overhead of VLA/WM-based autonomous control, especially when the channel coherence time is short. A promising direction is to exploit OTFS signaling to represent UAV channels in the delay-Doppler domain, where channel states are typically sparse, structured, and more stable under high mobility. Such compact representations may reduce state-update overhead and provide efficient inputs for VLM-based perception and WM-based prediction.}

\section{Conclusion}
This paper proposes an Embodied Agentic UAV framework for LAWNs, integrating VLA, WM, and Agentic AI to enable end-to-end closed-loop optimization from task semantics to executable control. The framework combines multimodal state perception, VLA-based decision-making, WM-driven environment simulation and policy refinement, and memory-reflection mechanisms for continual experience accumulation, while the embodied executor ensures stable and reliable hardware execution. Overall, it provides a systematic pathway for LAWNs to evolve toward predictive, actionable, and sustainably optimized autonomous networks.

\bibliographystyle{IEEEtran}
\bibliography{ref}
\section*{Biographies}

\textbf{Feibo Jiang} (jiangfb@hunnu.edu.cn) is currently an Associate Professor at Hunan Normal University, China.

\textbf{Li Dong} (Dlj2017@hunnu.edu.cn) is currently a Professor at Hunan University of Technology and Business, China.

\textbf{Lei Mao} (maolei@hunnu.edu.cn) is currently pursuing the master’s degree at Hunan Normal University, China.

\textbf{Kezhi Wang} (Kezhi.Wang@brunel.ac.uk) is a Professor with the Department of Computer Science, Brunel University London, U.K.

%\textbf{Kun Yang} (kunyang@nju.edu.cn) 
%received his PhD from the Department of Electronic \& Electrical Engineering of University College London (UCL), U.K. He 
%is currently a Chair Professor in the School of Intelligent Software and Engineering, Nanjing University, China.

\textbf{Cunhua Pan} (cpan@seu.edu.cn) is currently a full professor in Southeast University, China.

\textbf{Dong In Kim} (dongin@skku.edu) is a Distinguished Chair Professor at Sungkyunkwan University, South Korea.

\textbf{Naofal Al-Dhahir} (aldhahir@utdallas.edu) is a Professor at University of Texas at Dallas, USA.
\end{document}